\documentclass[superscriptaddress,pra,twocolumn,amsmath,amssymb]{revtex4}
\usepackage{graphicx}
\usepackage{dcolumn}
\usepackage{bm}

\begin{document}


\title{Experimental Quantum Process Discrimination}

\author{Anthony Laing}
\affiliation{Centre for Quantum Photonics, H. H. Wills Physics Laboratory \& Department of Electrical and Electronic Engineering, University of Bristol, Merchant Venturers Building, Woodland Road, Bristol, BS8 1UB, UK}
\author{Terry Rudolph}
\affiliation{QOLS, Blackett Laboratory, Imperial College London,
Prince Consort Road, London SW7 2BW, UK and
Institute for Mathematical Sciences, Imperial College London,
53 Exhibition Road, London SW7 2BW, UK}
\author{Jeremy L. O'Brien}
\affiliation{Centre for Quantum Photonics, H. H. Wills Physics Laboratory \& Department of Electrical and Electronic Engineering, University of Bristol, Merchant Venturers Building, Woodland Road, Bristol, BS8 1UB, UK}

\begin{abstract}%
Discrimination between unknown processes chosen from a finite set is experimentally shown to be possible even in the case of non-orthogonal processes. We demonstrate unambiguous deterministic quantum process discrimination (QPD) of non-orthogonal processes using properties of entanglement, additional known unitaries, or higher dimensional systems.  Single qubit measurement and unitary processes and multipartite unitaries (where the unitary acts non-separably across two distant locations) acting on photons are discriminated with a confidence of  $\geq97\%$ in all cases.
\end{abstract}

\maketitle

The indistinguishability of non-orthogonal quantum states lies at the heart of quantum mechanics---it underpins the fundamental challenge of quantum state discrimination \cite{chefles-2000,ru-prl-68-010301,co-nat-446-774} and has been harnessed as a resource in quantum technologies \cite{gi-rmp-74-145}. Perfect identification of an unknown quantum process that acts on the state of a quantum system (including unitary operations, measurements, and decohering processes) can be achieved via quantum process tomography, but requires infinite uses of the unknown process \cite{NielsenChuang}. Here we experimentally demonstrate that discriminating between non-orthogonal projective measurements and unitary operations can be achieved with finite uses of the unknown process, in stark contrast to the situation for quantum states. We use either entanglement or an additional known process to deterministically and unambiguously discriminate between non-orthogonal measurement processes, and qubit and qutrit unitary processes. Finally we experimentally demonstrate that non-local multipartite unitary processes can be locally distinguished---\emph{i.e.} without entanglement. Our processes act on photons and are discriminated with a confidence of  $\geq97\%$ in all cases.

Non-orthogonal quantum states cannot be distinguished with certainty because measurement of the state of a quantum system necessarily disturbs that state; to correctly identify a state chosen from a set of non-orthogonal states, one requires an infinite number of copies of the system prepared in the unknown state\cite{chefles-2000}.  Furthermore, the closer together (or less orthogonal) those states are that make up the set, the more difficult it is to tell them apart---\emph{i.e.} with a finite number of copies of the system prepared in the unknown state available to measure, the less certain of correct identification one can be.

\begin{figure}[b]
  \includegraphics[width=7.5cm]{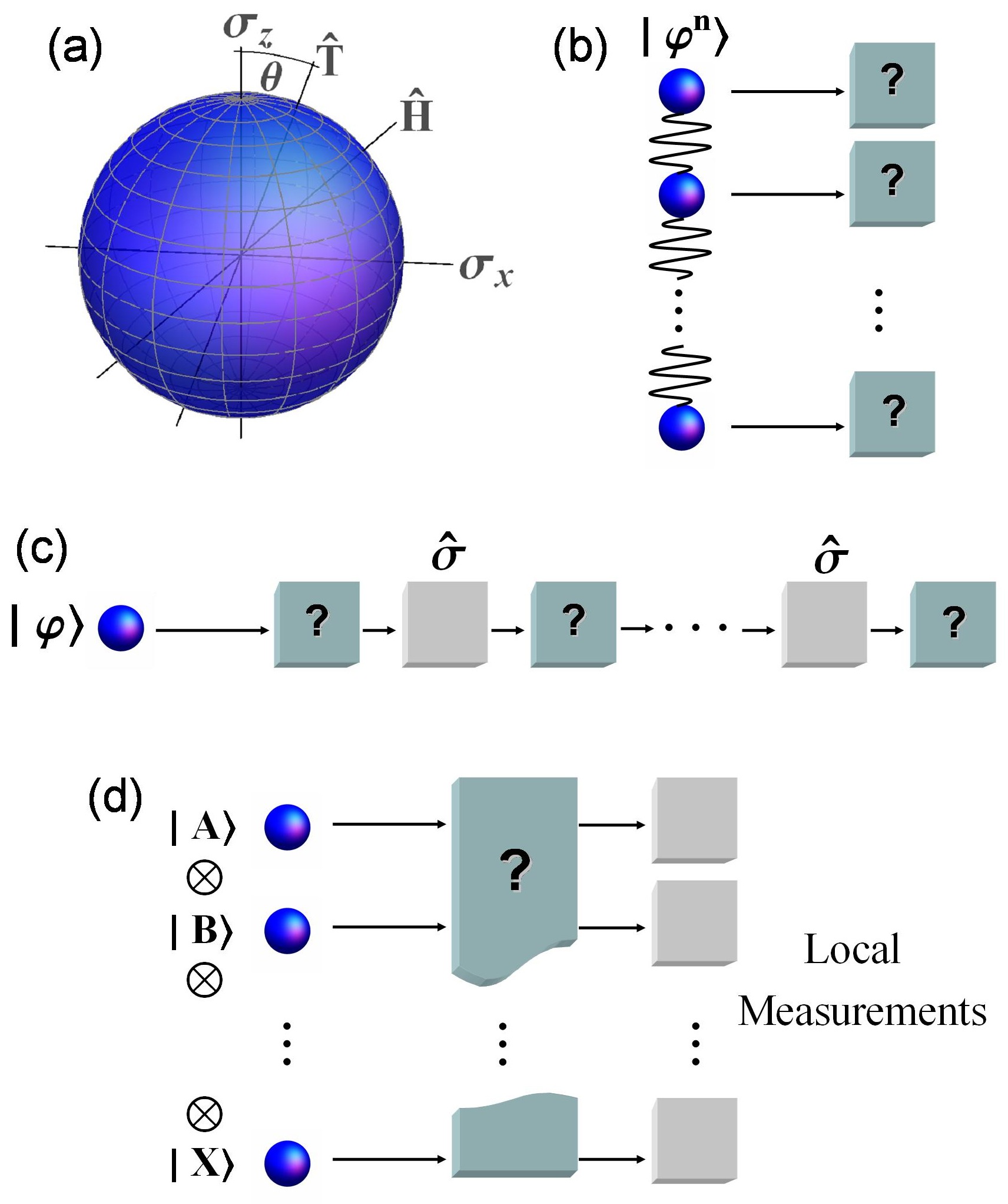}\\
  \caption{Experimental quantum process discrimination. (a)The Bloch Sphere showing nonorthogonal measurement directions $X$ and $Z$ and nonorthogonal Hermitian unitaries $\sigma_{z}$ and \^{H}.  $\sigma_{z}$ and \^{T} can always be discriminated with a finite number of qubits with $\theta$ arbitrarily acute. (b) Entanglement assisted QPD. (c) QPD without entanglement. (d) Multi-partite unitary discrimination without entanglement.}\label{Figure 1}
\end{figure}

\begin{figure*}[t]
  \includegraphics[width=15cm, height=7.85cm]{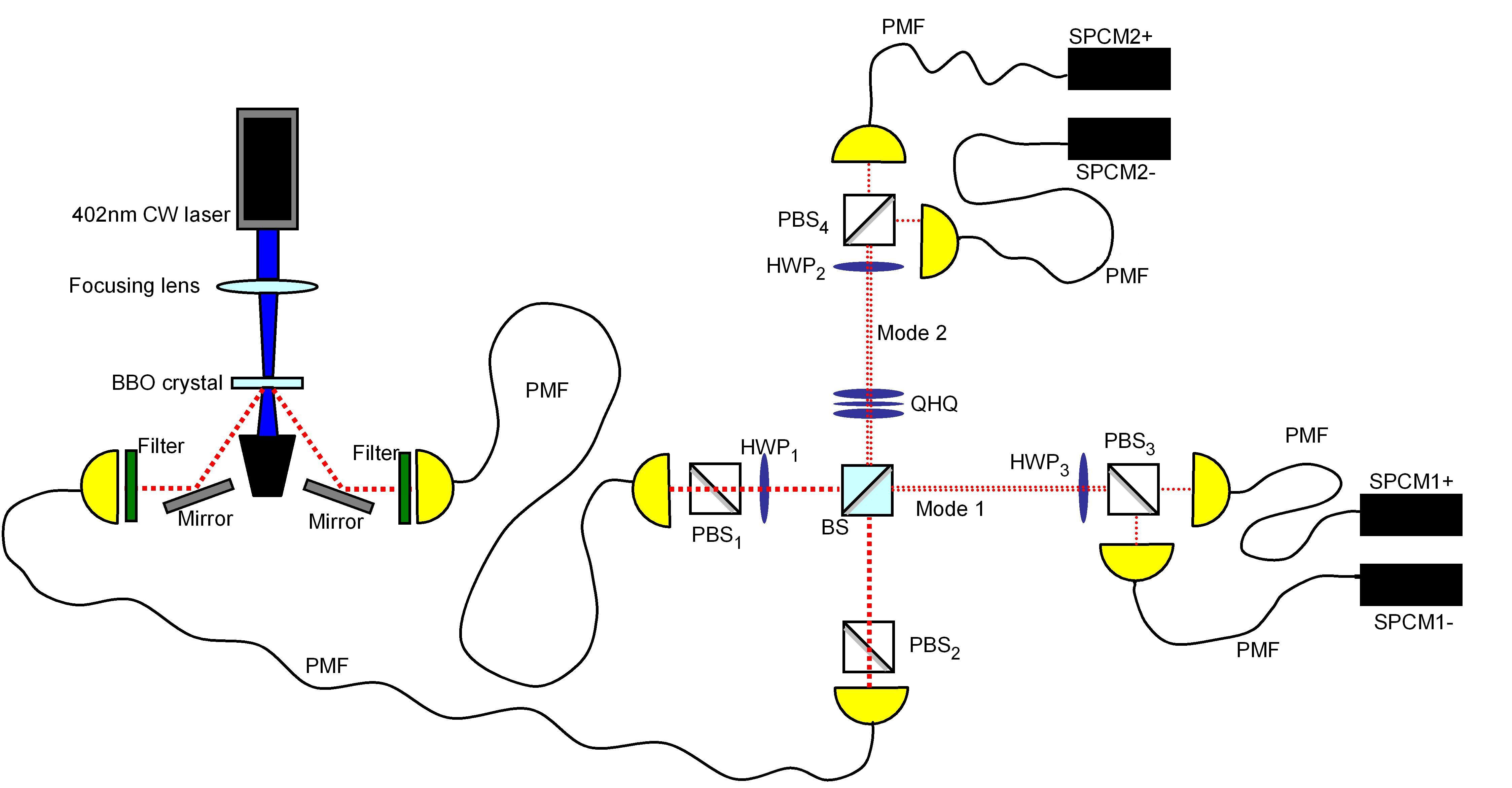}\\
  \caption{Experimental QPD.  A 2 mm Type 1 BBO crystal is pumped with a vertically polarised 60 mW 402 nm continuous wave laser. Pairs of horizontally polarized 804 nm photons are detected at a rate $\approx$ 3,500 Hz when collected into single mode polarisation maintaining fibres (PMF) after 2 nm interference filters.  Polarizing beamsplitters (PBS1 and PBS2) further purify the polarization; indistinguishability of the photons was confirmed by a Hong-Ou-Mandel dip \cite{ho-prl-59-2044} visibility of $96.9\pm0.3\%$.  A half-wave plate (HWP1) rotates the photon in mode 1 from horizontal (H) to vertical (V) and the two orthogonally polarised photons impinge on the input ports of the central 1/2 reflectivity beamsplitter (BS) to produce the output state: $|\psi_{out}\rangle = (|H_{1}V_{1}\rangle + |H_{1}V_{2}\rangle + |V_{1}H_{2}\rangle + |V_{2}H_{2}\rangle)/2$, where the subscript labels the mode.  Phase corrections are implemented with the quarter-half-quarter-wave plate combination (QHQ2) \cite{langfordthesisWavePlates} on mode 2. The optic axes of HWP2 and HWP3 are set to 0$^\circ$ or 22.5$^\circ$ to perform $\sigma_{z}$ or \^{H} respectively. Orthogonal polarisations are detected by a PBS followed by two single photon counting modules (SPCMs) in each mode. Each SPCM corresponds to a particular eigenvalue---\emph{eg} in mode 1, if we have set HWP3 to 22.5$^\circ$ to measure in the Bloch x-axis (D/A basis), a diagonally polarised photon would be transmitted to the SPCM corresponding to a `+1' eigenvalue, while an antidiagonal photon would be reflected to the SPCM corresponding to the `-1' eigenvalue.}\label{Figure 2}
 \vspace{-0.3cm}
 \end{figure*}

Quantum processes act to transform the state of a quantum system and include unitary operations, such as quantum logic gates; measurements, including von Neumann and more generalized POVMs; and decohering or dissipative processes. Any quantum process acting on a $d$ dimensional Hilbert space can be expressed as a quantum state in a $d^{2}$ Hilbert space \cite{NielsenChuang}. It may therefore seem natural to conclude that quantum process discrimination (QPD), where we wish to identify an unknown process chosen from a set of non-orthogonal processes,  is exactly analogous to quantum state discrimination. However, QPD is distinctly different to quantum state discrimination: in contrast to quantum states, quantum processes can be probed without disturbing the process itself; this expands the resources that can be used to tackle the problem---one can imagine using entangled states or additional known processes, for example. We show that these differences make QPD tractable in several examples of distinguishing non-orthogonal quantum measurements and unitary operations with a small number of applications of the unknown process.  These results generalize to allow arbitrarily close quantum processes to be distinguished with finite number of uses of the process. QPD promises fundamental insights into the nature of quantum mechanics as well as potential future applications to quantum technologies.

\emph{Entanglement-assisted measurement QPD.---}Our first experiment discriminates between non-orthogonal quantum measurements and harnesses the quantum correlations inherent in entangled states [Fig. 1(b)]. For example the Bell state $|\psi^{+}\rangle \equiv (|01\rangle + |10\rangle)/\sqrt{2}$ produces anti-correlated measurement outcomes when each qubit is measured along the Bloch z-axis; while expressed in the x-basis $|\psi^{+}\rangle = (|++\rangle - |--\rangle)/\sqrt{2}$ (where $|+\rangle\equiv|0\rangle+|1\rangle; |-\rangle\equiv|0\rangle-|1\rangle$), thereby resulting in perfect correlation when both qubits are measured along the Bloch x-axis (Fig. 1). In this way we can unambiguously and deterministically distinguish between a measurement along these two non-orthogonal (in Hilbert Space) axes with two uses of the process.

\begin{figure*}[t!]
  \includegraphics[width=17cm]{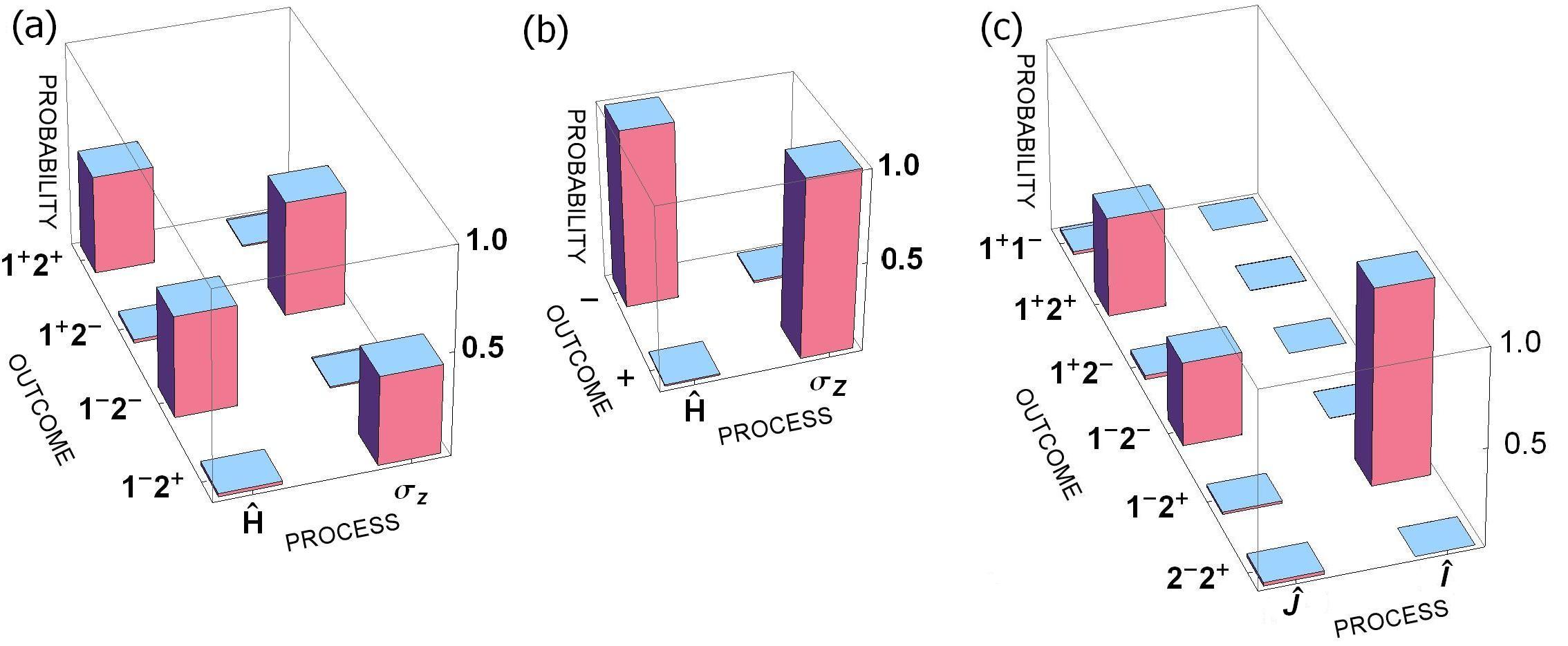}\\
  \caption{Experimental results for QPD of non-orthogonal quantum processes. (a) Entanglement assisted QPD of non-orthogonal measurement bases distinguishes between measurements in the H/V (along Bloch z-axis) and D/A basis (along Bloch x-axis) with a confidence of $(97.8 \pm 0.05)\%$.  Changing measurement basis from H/V to D/A is achieved by changing unitary from $\sigma_{z}$ (half-wave plate with optic axis set at 0$^\circ$) to \^{H} (half-wave plate with optic axis set at 22.5$^\circ$). (b) The scheme demonstrating unitary QPD without entanglement discriminates between $\sigma_{z}$ and \^{H} with a confidence of $(99.0 \pm 0.02)\%$. (c) The bipartite unitaries $\hat{J}$ and $\hat{I}$ are distinguished with a $(96.6 \pm 0.4)\%$ confidence.}\label{Figure 3}
  \vspace{-0.25cm}
\end{figure*}

Figure 2 shows the experimental set up used to demonstrate this QPD scheme.  The Bell state is encoded in the horizontal $H$ and vertical $V$ polarization of two photons in two spatial modes, 1 and 2: $|\psi^{+}\rangle = (|H_{1}V_{2}\rangle + |V_{1}H_{2}\rangle)/\sqrt{2}$, which is created by impinging an $H$ and $V$ polarised photon at the input ports of the 1/2 reflectivity beamsplitter \cite{ou-prl-61-50,ed-jjap-46-7175}. Ideally, when measuring in the H/V (Bloch z-axis) basis, only anti-correlations and no correlations should be observed (\emph{i.e.} +1,-1 or -1+1 eigenvalues will be measured); in contrast a measurement in the diagonal ($|D\rangle\equiv|$H$\rangle+|$V$\rangle$)/anti-diagonal ($|A\rangle\equiv|$H$\rangle-|$V$\rangle$) (Bloch x-axis) basis should give perfect correlations (+1+1 or -1-1).  Fig. 3(a) shows the results of this experiment;  to quantify, we are $97.8 \pm 0.05\%$ certain of a correct identification.  Note that this value is higher than the $96.9\%$ visibility of the Hong-Ou-Mandel Dip since a non-interfering state will make a correct identification half of the time. The resources used in this discrimination are one ebit of entanglement \cite{be-pra-53-2046} and two uses of the unknown device.

This measurement QPD scheme can be generalised to distinguish between any two single qubit observables, $\hat{S}$ and $\hat{T}$, which correspond to any arbitrary axes of the Bloch sphere [Fig. 1(a,b)]\cite{ji-prl-96-200401}.  Without loss of generality $\hat{S}$ can correspond to projections along the $z$ axis while $\hat{T}$ can be some other axis in the $x$-$z$ plane of the sphere (Fig. 1).  An n-qubit $W$ state ($W^{(n)}\equiv(|0...01\rangle+|0...10\rangle+...+|100...\rangle)/\sqrt{n}$) can be used to distinguish $\hat{S}$ and $\hat{T}$ if the angle between them is $\theta$ = 2 ArcTan$(1/\sqrt{n-1})$ (note that $|\Psi^+\rangle$ is the $n=2$ W state).  If the ``black box" is used to measure each of the n qubits and one of the measurement eigenvalues is $-1$ then the it is $\hat{S}$ 
 otherwise the black box is $\hat{T}$.  With an extra known measurement, one can choose any arbitrary $\theta$ between $\hat{S}$ and $\hat{T}$ and make $\theta$ as acute as one likes---the discrimination is always possible with a finite number of uses of the black box!

\emph{Entanglement-assisted unitary QPD.---}Entanglement is also useful for discrimination of non-orthogonal unitary processes. In particular, discrimination between the non-orthogonal single qubit unitaries Pauli-z ($\sigma_{z}$) and Hadamard \^{H} = $(\sigma_{z} + \sigma_{x})/\sqrt{2}$ is closely related to the measurement-QPD scheme above. The change of measurement basis in that scheme, between Bloch z-axis (H/V) and Bloch x-axis (D/A), is achieved experimentally with a half-wave plate (HWP) set to implement either a $\sigma_{z}$ or a \^{H}, respectively, followed by measurement in the H/V basis.  Applying \^{H} to each of the qubits of the $|\psi^{+}\rangle$ state, transforms it to the $|\phi^{-}\rangle = (|00\rangle - |11\rangle)/\sqrt{2}$ state, while the invariant effect of a $\sigma_{z}$ unitary applied to both qubits leaves the state unchanged.  From the results in Fig. 3(a) the discrimination is achieved with $(97.8 \pm 0.05)\%$ certainty. This unitary QPD requires 1 ebit of entanglement and two uses of the unknown process and can be generalised: It is always possible to discriminate between two unitary operations no matter how close together they are, with the use of entanglement and multiple, but finite, uses of the unknown unitary [Fig. 1(b)]\cite{ac-prl-87-177901, da-prl-87-270404}.  

We emphasize that the example we have demonstrated necessitated
utilizing entanglement and more than one application of the unknown
process. By contrast, were the unknown unitary operations $\sigma_{z}$ and $\sigma_{x}$, then
simply preparing a single qubit in the state $|0\rangle$ and making one use of
the process would have sufficed (the operations are orthogonal with
respect to the CB-norm \cite{cb-norm}). In fact the four processes
I, $\sigma_{x}$, $\sigma_{y}$, $\sigma_{z}$ are distinguishable either with just two unentangled qubits and
two uses of the unknown process, or by using entangled qubits and just
one use of the unknown process (as in superdense coding \cite{be-prl-69-2881}). By contrast we
have demonstrated distinguishing genuinely non-orthgonal operations with
a finite number of uses of the devices, a far more interesting scenario.

\emph{Unitary QPD without entanglement.---}Somewhat surprisingly, discrimination of non-orthogonal unitary operators can also be achieved using a single unentangled qubit with the aid of an extra known unitary [Fig. 1(c)].  For example $\sigma_{z}$ and $\hat{H}$ can be distinguished with the use of an additional $\sigma_{z}$: applying $\hat{U}$-$\sigma_{z}$-$\hat{U}$ to the $|0\rangle$ state retains the $|0\rangle$ state if the unknown unitary $\hat{U}$ happens to be $\sigma_{z}$, but produces the orthogonal $|1\rangle$ state if it is $\hat{H}$.  Clearly no entanglement is used here; the resources are two uses of the unknown device and one use of an additional known unitary.  We implement these single qubit unitaries using HWPs and the results are shown in Fig. 3(b). The certainty of distinguishing the non-orthogonal unitaries is $99.0 \pm 0.3\%$.  This scheme can be generalised to distinguish, without entanglement, any two non-orthogonal unitaries if the unknown unitary is complemented by particular known unitaries and the unknown device is used a multiple but finite number of times [Fig. 1(c)]\cite{du-prl-98-100503}.

\emph{Multipartite QPD without entanglement.---}Finally, we address the interesting case of bipartite unitary discrimination, where the unitary operator acts non-separably across two distinct subsystems, Alice's and Bob's say [Fig. 1(d)]. Remarkably, this type of discrimination is possible even in the case when Alice and Bob are only allowed to prepare their states locally (\emph{i.e.} they share no entanglement) and to perform local operations and classical communication (LOCC).

We experimentally demonstrate this interesting type of ``local QPD of non local processes''  by allowing Alice and Bob each access to one spatial mode with two polarization modes and 0, 1 or 2 photons. The state of Alice's system can be expressed in the Fock basis \{$|0\rangle_{A}\equiv|0\rangle; |1\rangle_{A} \equiv |1_{H}\rangle; |2\rangle_{A} \equiv |1_{V}\rangle; |3\rangle_{A} \equiv |2_{H}\rangle; |4\rangle_{A} \equiv|1_{H}, 1_{V}\rangle; |5\rangle_{A} \equiv |2_{V}\rangle$\}, and similarly Bob's. Our unitary operations across the combined Hilbert space are the identity $\hat{I}=I^{(6)}\otimes  I^{(6)}$, implemented by removing the beamsplitter and waveplates in Fig. 2, and $\hat{J}\equiv (H^{(6)}\otimes H^{(6)}).\hat{B}$, which is implemented by the beamsplitter $\hat{B}$ and HWP$_{2,3}$ set to implement
 \begin{equation}
 H^{(6)} = \left(
           \begin{array}{ccc}
             1 &
                   \begin{array}{cc}
                     0 & 0 \\
                   \end{array}
              &
                  \begin{array}{ccc}
                    0 & 0 & 0 \\
                  \end{array}
               \\
               \begin{array}{c}
                 0 \\
                 0 \\
               \end{array}
              & \hat{H} &
                      \begin{array}{ccc}
                        0 & 0 & 0 \\
                        0 & 0 & 0 \\
                      \end{array}
               \\
               \begin{array}{c}
                 0 \\
                 0 \\
                 0 \\
               \end{array}
              &
                  \begin{array}{cc}
                    0 & 0\\
                    0 & 0 \\
                    0 & 0 \\
                  \end{array}
               & H^{(3)} \\
           \end{array}
           \right)
\end{equation}
where
\begin{equation}
 H^{(3)} = \frac{1}{2}\left(
             \begin{array}{ccc}
               1 & \sqrt{2} & 1 \\
               \sqrt{2} & 0 & -\sqrt{2} \\
               1 & -\sqrt{2} & 1 \\
             \end{array}
           \right)
\end{equation}

By locally preparing the state $|\psi\rangle = |2 \rangle_{A}\otimes|1 \rangle_{B}$, we see that Alice and Bob can deterministically distinguish $\hat{J}$ from the identity operator $\hat{I}$.  Specifically, the input state $|\psi\rangle$ evolves under $\hat{J}$ to $\frac{1}{2\sqrt{2}}(|0 \rangle_{A}|3 \rangle_{B} - |0 \rangle_{A}|5 \rangle_{B} + |1 \rangle_{A}|1 \rangle_{B} - |2 \rangle_{A}|2 \rangle_{B} + |3 \rangle_{A}|0 \rangle_{B} - |5 \rangle_{A}|0 \rangle_{B})$ but is invariant under the identity $\hat{I}$. These states are orthogonal, and moreover locally distinguishable: in particular we  notice that a $1-$ and $2+$ outcome identifies the unknown operator as $\hat{I}$, whereas any $+/+$ or $-/-$ outcome identifies the operator as $\hat{J}$.  The measured data are shown in Fig. 3(d), demonstrating a discrimination confidence of $96.6 \pm 0.4\%$. Remarkably, this type of LOCC discrimination is always possible for all multi-partite unitaries [Fig. 1(d)] \cite{zh-prl-99-170401,du-prl-100-020503}.

\emph{Outlook---}One obvious experimental benchmark in QPD is the mutual non-orthogonality of the two processes that are discriminated.  It is natural to introduce an angle of distinguishability, which for single qubit processes is the angle between two axes of the Bloch sphere.  In the single qubit processes presented here, we have distinguished measurements separated on the Bloch sphere by an angle of 90$^\circ$  and unitaries separated by a Bloch sphere angle of 45$^\circ$. In the case of non-deterministic QPD \cite{zi-arXiv:0712.3675} the success probability must be considered.

The dynamics of the world around us, and our interactions with it, all amount to quantum processes. It is quite remarkable how different is our ability to probe processes using known quantum states from our ability to probe states using known processes. It seems clear that understanding this asymmetry has implications for our foundational understanding of quantum mechanics as well as for the pragmatic considerations which underscore the emerging potential of quantum information science. Connections between QPD and quantum cloning \cite{da-pra-67-042306,br-pra-62-012302} may exist.

We would like to thank S. D. Bartlett, A. S. Clark, G. Gabaston, B. P. Lanyon, J. C. F. Matthews, A. Politi, S. Popescu, J. G. Rarity, T. Short, A. Stefanov and A. B. Young. This work was supported by the US Disruptive Technologies Office (DTO), the UK Engineering and Physical Sciences Research Council (EPSRC), the UK Quantum Information Processing Interdisciplinary Collaboration (QIP IRC),  and the Leverhulme Trust.

\end{document}